\documentclass{aa} 

\usepackage{graphicx}
\usepackage{lscape}
\usepackage{subfigure}
\usepackage{natbib}
\usepackage{hyperref}
\usepackage[varg]{txfonts}
\usepackage{longtable}
\usepackage{booktabs}
\usepackage[normalem]{ulem} 
 \usepackage{float}
\usepackage{multirow}
\usepackage{soul}
\usepackage[table,x11names]{xcolor}
\usepackage{scalerel}
\usepackage{tabularx}
\usepackage{tikz}
\usepackage[T1]{fontenc}
\usepackage{needspace}

\usetikzlibrary{svg.path}

\definecolor{orcidlogocol}{HTML}{A6CE39}
\tikzset{
  orcidlogo/.pic={
    \fill[orcidlogocol] svg{M256,128c0,70.7-57.3,128-128,128C57.3,256,0,198.7,0,128C0,57.3,57.3,0,128,0C198.7,0,256,57.3,256,128z};
    \fill[white] svg{M86.3,186.2H70.9V79.1h15.4v48.4V186.2z}
                 svg{M108.9,79.1h41.6c39.6,0,57,28.3,57,53.6c0,27.5-21.5,53.6-56.8,53.6h-41.8V79.1z M124.3,172.4h24.5c34.9,0,42.9-26.5,42.9-39.7c0-21.5-13.7-39.7-43.7-39.7h-23.7V172.4z}
                 svg{M88.7,56.8c0,5.5-4.5,10.1-10.1,10.1c-5.6,0-10.1-4.6-10.1-10.1c0-5.6,4.5-10.1,10.1-10.1C84.2,46.7,88.7,51.3,88.7,56.8z};
  }
}

\newcommand\orcidicon[1]{\href{https://orcid.org/#1}{\mbox{\scalerel*{
\begin{tikzpicture}[yscale=-1,transform shape]
\pic{orcidlogo};
\end{tikzpicture}
}{|}}}}
\bibpunct{(}{)}{;}{a}{}{,} 

\usepackage{amstext}
\usepackage{amsmath}
\hypersetup{
    colorlinks=true,
    citecolor=blue,
    linkcolor=blue,
    urlcolor=blue,
    }

\makeatletter
\renewcommand*\aa@pageof{, page \thepage{} of \pageref*{LastPage}}
\makeatother
\definecolor{imputedR}{rgb}{0.99608, 0.87843, 0.56471}
\definecolor{imputedG}{rgb}{1, 1, 0.74902}
\definecolor{both}{rgb}{0.56863, 0.74902, 0.85882}
\definecolor{cornflowerblue}{rgb}{0.39, 0.58, 0.93}
\definecolor{ste}{rgb}{0.99, 0.00, 0.5}

\definecolor{mpc}{rgb}{0.0, 0.3, 0.7}
\definecolor{mpc2}{rgb}{0.4, 0.1, 0.5}

\begin{document}

   \title{Scavenger hunt: Selection of obscured active galactic nuclei combining multiband optical variability and colors\thanks{Observations were provided by the ESO programs 088.D-4013, 092.D-0370, and 094.D-0417 (PI G. Pignata).}}
   \titlerunning{Scavenger hunt: selection of obscured AGN combining multiband optical variability and colors}

   \author{D. De Cicco\inst{1,2,\orcidicon{0000-0001-7208-5101}}, S. Cavuoti\inst{2,3,\orcidicon{0000-0002-3787-4196}}, M. Paolillo\inst{1,2,3,\orcidicon{0000-0003-4210-7693}}, V. Petrecca\inst{3,\orcidicon{0000-0002-3078-856X}}, Y. Maruccia\inst{3,\orcidicon{0000-0003-1975-6310}}, P.~Sánchez-Sáez\inst{4,\orcidicon{0000-0003-0820-4692}}}
   \authorrunning{D. De Cicco et al.}

\institute{
Department of Physics, University of Napoli ``Federico II'', via Cinthia 9, 80126 Napoli, Italy
\\e-mail: demetra.decicco@unina.it
\and
INAF - Osservatorio Astronomico di Capodimonte, via Moiariello 16, 80131 Napoli, Italy 
\and
INFN - Sezione di Napoli, via Cinthia 9, 80126 Napoli, Italy 
\and 
European Southern Observatory, Karl-Schwarzschild-Strasse 2, 85748 Garching bei M\"{u}nchen, Germany
  }

\date{}
  \abstract
   {
   As wide-field optical surveys such as Vera C. Rubin Observatory’s Legacy Survey of Space and Time (LSST) begin operations, time-domain astronomy is about to face a data revolution, paving the road for new and expanded variability studies.
   }
   {
   This work leverages the complementary power of optical variability and color selection to identify active galactic nuclei (AGN), with a particular emphasis on optimizing the identification of obscured AGN, typically more challenging to distinguish from inactive galaxies based on optical variability alone. The analysis is designed to provide valuable insights in the context of performance forecasting for the LSST, albeit using a scaled-down version of the LSST dataset.
   }
   {
   We present the first combined AGN selection based on $g$-, $r$-, and $i$-band light curves from the VST-COSMOS survey, spanning a 3.3 yr baseline. We identify AGN candidates independently in each band using a random forest (RF) classifier trained on features primarily related to optical variability, along with six optical/infrared colors and a morphology indicator. We subsequently merge the resulting band-specific samples in order to enhance selection purity and reliability. We then focus on defining a subset of features that significantly improve the identification of obscured AGN.
    }
   { 
   The RF classifiers yield a consistent performance across the three bands, highlighting the critical role played by contamination. Using the combined three-band plus color selection we successfully recover $58^{+9}_{-8}\%$ of all AGN and $69^{+10}_{-8}\%$ of the known obscured AGN that have been independently confirmed in all three bands. When requiring confirmation in at least two out of the three bands, these fractions increase to $69^{+10}_{-8}\%$ and $80^{+10}_{-9}\%$, respectively. Moreover, we demonstrate that, while combining variability features with colors is crucial to improve obscured AGN selection, relying solely on color features results in a markedly higher contamination rate.
   }
   {}

   \keywords{}

   \maketitle

\section{Introduction}
\label{section:introduction}
Active galaxies are among the most luminous and variable extragalactic sources, powered by accretion onto supermassive black holes (SMBHs) hosted in their centers. Optical variability, generally detected on timescales from days to years, has proven to be a powerful tool for the identification of unobscured (or Type I) active galactic nuclei (AGN) in time-domain photometric surveys. Variations are aperiodic and trace instabilities in the accretion flow, probing the physical mechanisms at play in the surroundings of the SMBH.

However, the situation is markedly different for obscured (or Type II) AGN. In the scenario of the unified model \citep{Antonucci,Urry&Padovani}, the lack of broad optical emission lines in this class of sources is explained by the presence of a dusty torus preventing us from directly observing the accretion disk and broad-line region. As a result, the central engine is largely hidden in the optical, and what we detect is mostly emission from the narrow-line region and the host galaxy, with minimal contribution from the variable nucleus. Optical variability is therefore expected to be much weaker or absent \citep{lopeznavas}, posing significant challenges for the selection of obscured AGN based on such a property. 
Despite theoretical predictions, a growing number of obscured AGN have been found to exhibit optical variability \citep[e.g.,][]{lopeznavas22,bernal25}, but the effectiveness of optical variability as a selection tool for obscured AGN remains a matter of active debate (see for instance the recent review from \citealt{PP2025}). This is intriguing, especially in the framework of upcoming large-scale multivisit surveys, such as the Vera C. Rubin Observatory's Legacy Survey of Space and Time (LSST; \citealt{ivezich19}), expected to deliver vast amounts of high-cadence photometric data. Understanding the mechanisms behind optical variability of obscured AGN is therefore critical to improve AGN selection techniques, as well as to gain a deeper insight into AGN general structure and evolution.

This work is part of a series focused on AGN selection in the COSMOS field, based on the analysis of optical variability from time series obtained via the VLT Survey Telescope (VST; \citealt{VST}): this is a 2.6-meter optical telescope, with a field of view of 1 sq. deg. and a pixel scale of 0.214\arcsec; it is located in Chile, at the Paranal Observatory, and is dedicated to wide-field imaging surveys of the southern sky \citep[see, e.g.,][]{Cappellaro15,Falocco15,botticella17,Fu18,Liu18,Liu20,Poulain20}.
Our dataset covers a 3.3 yr baseline and consists of observations in the $g$, $r$, and $i$ bands; while we made extensive use of $r$-band data in \citet{decicco15,decicco19,decicco21,decicco22,decicco25,cavuoti24,maruccia25,kaviraj26}, the other two datasets are here used as a whole for the first time for this kind of analysis, though part of the $g$-band data were used in \citet{Lira24,decicco25}. 

The VST-COSMOS dataset can be regarded as a scaled-down analog of what the LSST is expected to deliver. While the analysis presented here is limited to a 3.3 yr baseline, our current monitoring extends beyond 11 yr, i.e., longer than the 10 yr baseline planned for LSST. Moreover, the single-visit depth of VST and LSST images are comparable. On the other hand, the VST-COSMOS area covers roughly one tenth of the area the LSST survey will cover per deep drilling field (DDF), and our analysis is limited to three optical bands, versus the six that LSST will provide. Given these considerations, one of the goals of our series of works has been to provide lower-limit expectations for the AGN selection performance of the LSST. 

The primary goal of this work is to independently exploit the three optical bands available from the VST-COSMOS dataset and integrate each with static colors and morphological information from the literature, thereby combining the diagnostic power of optical variability and color selection to enhance AGN identification and expand the census of obscured AGN beyond that achieved in earlier studies.
In particular, this work builds upon the approach of \citet{decicco25}, who made use of a random forest (RF; \citealt{Breiman2001}) algorithm to identify AGN by testing different sets of features, and also focused on optimizing the selection of obscured AGN combining $r$- and $g$-band features together, identifying a feature set that returned a ($68.1 \pm 1.2$)\% completeness for this class of sources. While this number is significantly lower than the near-100\% completeness typically obtained for bright, unobscured AGN, it still represents a great improvement compared to earlier results trying to detect the bulk of the obscured AGN population (see, e.g., \citealt{decicco19}). 
Here we make use of the same algorithm, but extend the analysis to three bands and expand the AGN labeled set (LS) used in the past encompassing sources with a wider diversity of properties. We propose a different approach, identifying AGN based on an independent selection across the three available bands. We then adopt the feature-optimization approach of \citet{decicco25} to identify the most effective predictors to unearth obscured AGN, but this time we do it for each band separately in order to obtain independent results. We then investigate the advantages of combining the obtained samples to enhance the purity of our AGN selection.

The structure of this paper is as follows. Section \ref{section:dataset} presents the dataset, the derived source catalogs, and the LS. Section \ref{section:classification} outlines the classification procedure carried out in the three bands, using two different LSs. Section \ref{section:colors} is dedicated to models using variability features or colors separately. Section \ref{section:agn2} addresses the optimization of obscured AGN selection. Finally, Section \ref{section:conclusions} summarizes our main findings. We note that throughout this work we generally use the terms ``obscured'' for AGN that were classified as Type II via optical spectroscopy as they are characterized by peculiar optical variability, as is detailed in Section \ref{section:agn2}.

\section{The VST-COSMOS \emph{gri} dataset}
\label{section:dataset}
As was mentioned above, our dataset consists of $g$-, $r$-, and $i$-band data. The $r$-band dataset consists of 54 visits, while the $g$- and $i$-band datasets have a lower observing cadence and count 25 and 23 visits, respectively; this is due to differences in the original cadence planning of the SUDARE-VOICE surveys \citep{Cappellaro15,Vaccari,botticella17}, designed for supernova studies, where these data come from.
The single-visit depth of the $g$-, $r$-, and $i$-band data is of 25.0, 24.6, and 24.0 mag, respectively, at a confidence level of $\sim5\sigma$. Each image originally covered an area of 1 sq. deg., but we masked $\approx17\%$ of the field in order to exclude saturated stars, defected areas, and the noisiest regions on the edges of each image.

Table \ref{tab:dataset} in Appendix \ref{appendixA} reports information about the three-band datasets. For general information about the data structure and the reduction process, we refer the reader to Section 2 of \citet{decicco15} and Section 2 of \citet{decicco19}.

\subsection{The \emph{gri} catalogs}
\label{section:catalogs}
The catalog extraction and aperture correction processes follow what was done for the $r$-band data in \citet{decicco19}, described in detail in Section 3 of that work. In short, we used SExtractor \citep{bertin} to obtain from each visit in each band a catalog of sources. Consistent with \citet{decicco15,decicco19}, we chose to work with magnitudes measured within a 2\arcsec-diameter aperture, which typically encloses $\approx70\%$ of the flux from a point-like source. In order to correct for the effect of having different seeing values from visit to visit and to compensate for potential calibration offsets, we identified a reference visit for each band and normalized all the others to this. Specifically, the reference visits have the lowest possible seeing compatible with the absence of significant defects; we chose images \#48\footnote{We note that the reference visit for the $r$ band is the same we used in \citet{decicco19}. As in that work we excluded one visit but kept track of its ID number, the reference image was there identified as \#49. Here we did not keep track of that in order not to have missing numbers in the list. Indeed, the whole $r$-band dataset is a subset of the one we used in \citet{decicco19}.}, 48, and 29 from Table \ref{tab:dataset} for the $r$, $g$, and $i$ band, respectively (in bold in the table). 

We obtained a corrective factor for each visit by computing the average magnitude difference with respect to the reference image for all the sources with a magnitude in the range $16-21$~mag. We chose this magnitude range as it minimizes the influence of very bright or very faint objects. In our previous works we masked part of the $r$-band images (saturated stars, halos, defected areas, and the image edges) making use of the masks obtained via the \emph{Pulecenella} code \citep{Huang}. In \citet{decicco15,decicco19} we used the catalog obtained from a stacked image\footnote{The stacked image was produced as the median of all the exposures having a seeing full width at half-maximum (FWHM) < 0.80\arcsec; it has a total exposure time of 19800 s, and the limiting magnitude at $5\sigma$ above the background r.m.s. is r(AB) $\approx 26$ mag.} as a reference, and matched all the single-visit catalogs with that, obtaining a master catalog of 22927 sources, which we required being detected in at least half of the 54 $r$-band visits. Given the higher observing cadence of our $r$ band with respect to the other two, here we used that same $r$-band master catalog as a starting point for cross-matching with the $g$- and $i$-band catalogs. As a consequence, we did not need to apply masks to the $g$- and $i$-band images, as we simply cross-matched the corresponding master catalogs with the $r$-band one. Consistent with what we did for the $r$-band master catalog, we required that the $g$- and $i$-band master catalogs include sources detected in at least half the visits. We obtained 23990 and 26682 sources detected in at least 13 and 12 visits in the $g$ and $i$ band, respectively. The intersection of the three master catalogs returned a catalog of 16910 sources.

\subsection{The feature set for AGN classification}
\label{section:features}
This work is based on the use of the same features used in \citet{decicco19}, which we report in Table \ref{tab:features} for the sake of convenience. The list includes 29 variability features, one morphology feature, and six colors. 
The variability features have been widely used in the literature \citep[e.g.;][]{ponti12,Kim2016,CabreraVives2017,decicco21,PSS21} for time series analysis and are computed from the source light curves; we computed these features independently for each band. Given the different sampling in each band, the maximum number of points per light curve will differ from band to band. Nevertheless, since we used these features independently in each band, this is not an issue. The morphology feature comes from the COSMOS Advanced Camera for Surveys (ACS) catalog \citep{Koekemoer,scoville07b} from the \emph{Hubble} Space Telescope, based on F814W imaging. This feature is a neural network–based classifier that distinguishes stars from galaxies and separates extended from point-like sources based on their morphology. In light of our performance forecasting objectives, we chose colors derived from the $uBrizy$ filters, to mirror those that will be available from the LSST\footnote{Since $g$-band magnitudes are not provided in the chosen reference catalog (see further), we replaced them with the Subaru B band, which closely overlaps the $g$-band wavelength coverage.}. We also included a mid-infrared (MIR) color that existing or forthcoming ancillary datasets provide or will provide in the very near future. This color proved to be essential in distinguishing AGN from inactive galaxies, as is shown, for example, in \citet{decicco21}. The static color features were obtained from the corresponding magnitudes reported in the COSMOS2015 catalog \citep{laigle} and thus they were computed only once, and the same values for each source were used for classification in each band. Additional information about these features can be found in Sects. 2.2, 2.3, and 2.4 of \citet{decicco21}.

\begin{table*}[tb]
\caption{List of variability, morphology, and color features used in this work.}
\label{tab:features}      
 \renewcommand\arraystretch{1.2}
 \footnotesize
 \resizebox{\textwidth}{!}{
 \begin{tabular}{c l l l}
\toprule 
\ & Feature & Description & Reference\\
\hline
\multirow{9}{*}{{\rotatebox[origin=c]{90}{classic var. features}}} & \texttt{A$_{SF}$} & rms magnitude difference of the SF, computed over a 1 yr timescale & \citet{Schmidt}\\
& \ \texttt{$\gamma_{SF}$} & Logarithmic gradient of the mean change in magnitude & \citet{Schmidt}\\
& \ \texttt{GP\_DRW\_$\tau$} & Relaxation time $\tau$ (i.e., time necessary for the time series to become uncorrelated), & \citet{graham17}\\
& \  & from a DRW model for the light curve & \\
& \ \texttt{GP\_DRW\_$\sigma$} & Variability of the time series at short timescales, & \citet{graham17}\\
& \ & from a DRW model for the light curve & \\
& \ \texttt{ExcessVar} & Measure of the intrinsic light curve variance & \citet{Nandra1997,allevato}\\
& \ \texttt{P$_{var}$} & Probability that the source is intrinsically variable & \citet{mclaughlin}\\
& \ \texttt{IAR$_\phi$} & Level of autocorrelation using a  discrete-time representation of a DRW model & \citet{eyheramendy18}\\
\hline
\multirow{27}{*}{{\rotatebox[origin=c]{90}{FATS features}}} &\ \texttt{Amplitude} & Half of the difference between the median of the maximum 5\% and of the minimum & \citet{richards11}\\
 &\ & 5\% magnitudes & \\
 & \ \texttt{AndersonDarling} & Test of whether a sample of data comes from a population with a specific distribution & \citet{nun}\\
 & \ \texttt{Autocor\_length} & Lag value where the autocorrelation function becomes smaller than $\eta^e$ & \citet{kim11}\\
 & \ \texttt{Beyond1Std} & Percentage of points with photometric mag that lie beyond 1$\sigma$ from the mean & \citet{richards11}\\
 & \ \texttt{$\eta^e$} & Ratio of the mean of the squares of successive mag differences to the variance & \citet{kim14}\\
 & \ & of the light curve & \\
 & \ \texttt{Gskew} & Median-based measure of the skew & -\\
 & \ \texttt{LinearTrend} & Slope of a linear fit to the light curve & \citet{richards11}\\
 & \ \texttt{MaxSlope} & Maximum absolute magnitude slope between two consecutive observations & \citet{richards11}\\
 & \ \texttt{Meanvariance} & Ratio of the standard deviation to the mean magnitude & \citet{nun}\\
 & \ \texttt{MedianAbsDev} & Median discrepancy of the data from the median data & \citet{richards11}\\
 & \ \texttt{MedianBRP} & Fraction of photometric points within amplitude/10 of the median mag & \citet{richards11}\\
 & \ \texttt{MHAOV\_Period} & Period obtained via the Multi-Harmonic Analysis Of Variability periodogram & \citet{Huijse18}\\
 & \ \texttt{PairSlopeTrend} & Fraction of increasing first differences minus fraction of decreasing first differences & \citet{richards11}\\
 & \ & over the last 30 time-sorted mag measures & \\
 & \ \texttt{PercentAmplitude} & Largest percentage difference between either max or min mag and median mag & \citet{richards11}\\
 & \ \texttt{Q31} & Difference between the 3\textsuperscript{rd} and the 1\textsuperscript{st} quartile of the light curve & \citet{kim14}\\
 & \ \texttt{Period\_fit} & False-alarm probability of the largest periodogram value obtained with Lomb-Scargle & \citet{kim11}\\
 & \ \texttt{$\Psi_{CS}$} & Range of a cumulative sum applied to the phase-folded light curve & \citet{kim11}\\
 & \ \texttt{$\Psi_\eta$} & $\eta^e$ index calculated from the folded light curve & \citet{kim14}\\
 & \ \texttt{R$_{cs}$} & Range of a cumulative sum & \citet{kim11}\\
 & \ \texttt{Skew} & Skewness measure & \citet{richards11}\\
 & \ \texttt{Std} & Standard deviation of the light curve & \citet{nun}\\
 & \ \texttt{StetsonK} & Robust kurtosis measure & \citet{kim11}\\

\hline
\multirow{4}{*}{{\rotatebox[origin=c]{90}{morph.}}}\  \\
& \texttt{class\_star} & \emph{HST} stellarity index & \citet{Koekemoer},\\
\ & &  & \citet{scoville}\\ \\
\hline
\multirow{6}{*}{{\rotatebox[origin=c]{90}{colors}}} & \ \texttt{u-B} & CFHT $u$ mag -- Subaru $B$ mag & \citet{laigle}\\
 & \ \texttt{B-r} & Subaru SuprimeCam $B$ mag -- Subaru SuprimeCam $r$+ mag & \citet{laigle}\\
 & \ \texttt{r-i} & Subaru SuprimeCam $r+$ mag -- Subaru SuprimeCam $i+$ mag & \citet{laigle}\\
 & \ \texttt{i-z} & Subaru SuprimeCam $i+$ mag -- Subaru SuprimeCam $z++$ mag & \citet{laigle}\\
 & \ \texttt{z-y} & Subaru SuprimeCam $z++$ mag -- Subaru Hyper-SuprimeCam $y$ mag & \citet{laigle}\\
\cline{2-4}
 & \ \texttt{ch21} & \emph{Spitzer} 4.5 $\mu$m (\emph{channel2}) mag -- 3.6 $\mu$m (\emph{channel1}) mag & \citet{laigle}\\
\bottomrule
\vspace{1mm}
\end{tabular}
}
\footnotesize{\textbf{Notes.} The first two sections of the table list the variability features used; \texttt{class\_star}, the only morphology feature used, follows; the bottom section reports the color features used: all of them are optical/near-infrared colors except for \texttt{ch21}, which is a MIR color. This table corresponds to Table 1 in \citet{decicco21}.}
\end{table*}

\subsection{Labeled set}
\label{section:LS}
As was mentioned in the introduction, this work builds on \citet{decicco21,decicco25}. Specifically, both works made use of a LS consisting of AGN with X-ray emission and selected via optical spectroscopy from the \emph{Chandra}-COSMOS Legacy Catalog\footnote{We stress that the AGN from the \emph{Chandra}-COSMOS Legacy Catalog have, by construction, X-ray detection, but were classified as Type I or Type II via optical spectroscopy.} \citep{marchesi} or AGN selected via their MIR properties based on the criterion by \citet{donley}, and of non-AGN which can be either stars or inactive galaxies, i.e., galaxies not showing any sign of nuclear activity. There are small differences between the LSs in \citet{decicco21} and \citet{decicco25} due to constraints originating from specific requirements. 

The present work tests the expansion of the AGN LS used in our previous works, and hence begins with two parts where we made use of two different LSs: the first one (hereafter, the spec-MIR LS) takes the cue from the one used in the two works mentioned above, but it turns out to be smaller as we require each source in the LS to be present in the $g$, $r$, and $i$ master catalogs. It includes 2154 sources, specifically:
\begin{itemize}
    \item 358 AGN (versus the 380 AGN used in in \citealt{decicco25}), of which 209 spectroscopically classified as Type I and 96 spectroscopically classified as Type II; 203 of these 358 sources are classified as AGN via their MIR properties following \citet{donley}, and 53 of these 203 sources do not have spectroscopic classification (MIR-only AGN subsample); 
    \item 1796 non-AGN (versus the 2163 non-AGN used in \citealt{decicco25}).
\end{itemize}

In the second part of the work we include in the LS another sub-class of AGN, i.e., 162 additional AGN selected only on the basis of their X-ray-to-optical flux ratio $X/O$ (following \citealt{Maccacaro}, but also \citealt{hornschemeier,Xue,civano}, we require them to have $X/O > -1$). Hereafter, we will label this subsample of 162 sources as X-ray-only AGN. The total LS in this second part (hereafter, the main LS) therefore consists of 2316 sources in total --1796 non-AGN and 520 AGN-- corresponding to an increase of nearly $50\%$ with respect to the AGN in the spec-MIR LS. The inclusion of X-ray-selected AGN introduces greater diversity, including also some low-luminosity AGN. We caution that the spec-MIR LS does include AGN also confirmed via X-ray properties, but we did not select them based on this property; hence they are always classified as AGN via at least one additional property, contrary to the X-ray-only AGN subsample that we added to the main LS. 

We stress that, consistent with previous works in this series, our classifiers are always binary, which means that we did not deal with sub-classification within the two classes of AGN and non-AGN, though we used that information when discussing the selection of obscured AGN. We also point out that color and morphological information is available for all the sources in our LS. Additional details about how we selected the samples of sources making up our LS are reported in Section 2.5 of \citet{decicco21} and Section 2.2 of \citet{decicco25}; hence we do not include them here for the sake of conciseness.

Table \ref{tab:LS} reports the number of sources belonging to AGN and non-AGN as well as to the various sub-samples of sources they consist of. We point out that spec-MIR LS is a subsample of the main LS, and also note that a source can be classified as AGN based on more than one diagnostic, which means that the sum of the numbers reported for the AGN class will exceed the corresponding number of AGN. 

\begin{table}[ht]
    \caption{Information about the number of sources in the LS and its various parts.}
    \renewcommand{\arraystretch}{1.3}
    \resizebox{\columnwidth}{!}{
    \begin{tabular}{lcccccc}
    \toprule
    \multirow{2}{*}{} & \multicolumn{4}{c}{AGN} & \multicolumn{2}{c}{non-AGN} \\
    \cmidrule(lr){2-5} \cmidrule(lr){6-7}
    \ & Type I & Type II & MIR & X-ray & Stars & Gal \\
    \midrule
    \ spec-MIR LS (358 AGN) & 209 & 96 & 203 & -   & 841 & 955 \\
   c\ &                            &     &  (53 MIR-only) & & &\\
    \hline
    \ main LS (520 AGN) & 209 & 96 & 203 & 477 & 841 & 955 \\
    \ &                            &     &  (53 MIR-only) & (162 X-ray only) & &\\
   \bottomrule
    \end{tabular}
    }
    \label{tab:LS}
    \footnotesize{\textbf{Notes.} The table shows the composition of the LS for the two parts of this work: the top line refers to the LS where we did not include AGN selected via their $X/O$, while these sources are included in the bottom line. The rest of the LS is unchanged and, in particular, the number of non-AGN in each LS is always 1796. The various columns indicate (left to right): AGN classified via optical spectroscopy as Type I or Type II, AGN selected via their MIR properties, AGN selected via their $X/O$, plus stars and inactive galaxies which make up the non-AGN part of the LS.}
\end{table}

\section{Classification via a random forest algorithm}
\label{section:classification}
This work aims at combining AGN classification in three different bands, and starts with two separate experiments based on the use of two distinct LSs, as is detailed in Table \ref{tab:LS}. In both parts we focused on the classification of the sources in our LS via a RF algorithm. The classification algorithm was used on the $g$, $r$, and $i$ bands independently, then we also cross-matched the obtained results. Color features and a morphology indicator are always added to the list of variability-based features used for each band.

Our code is based on the use of the Python \emph{scikit-learn} library \citep{scikit-learn}. In what follows we essentially adopted the same approach described in Section 4 of \citet{decicco25}, which here we briefly summarize. Since our LS is unbalanced, in the algorithm we set \texttt{class\_weight = balanced\_subsample}: in this way, every time a bootstrap sample is used to extract a subset of features to build a tree, the algorithm adjusts the weights of each class dynamically based on the class distribution in that specific bootstrap sample used to train the corresponding tree. 

We optimized the performance of each RF classifier, operating on the two possible LSs in each band, testing a set of possible combinations of the main hyperparameters involved in the RF building process\footnote{In machine learning, the settings you choose before training an algorithm are called hyperparameters, and they define how the algorithm will work. For a RF algorithm they will establish, for instance, the number of trees to be used in the forest, the maximum depth of each tree, the number of features to consider when splitting a node.}. This was achieved via a grid search, which explores all the combinations of the selected hyperparameter values using cross-validation and predefined scoring metrics to identify the optimal set. 
Given the potentially infinite number of possible combinations and the high computational cost, it was necessary to constrain the set of values explored for each hyperparameter. We tested 216 combinations of hyperparameter per classifier, focusing on the following five hyperparameters: \textit{n\_estimators}, \textit{min\_samples\_split}, \textit{max\_depth}, \textit{min\_samples\_leaf}, \textit{max\_features}.
In Appendix \ref{appendixB} we provide a short description of these five hyperparameters, together with details about the tested values and the obtained best sets. We evaluated model performance via the balanced accuracy, which averages the recall of both classes and accounts for the class imbalance in our LS. 

A common approach in machine learning is to divide the LS into two disjoint subsets: a training set (typically 70–75\%) for model fitting, and a validation set (30–25\%) for performance evaluation during training. However, due to the heterogeneous nature of our AGN sample -- selected on the basis of different properties -- this strategy would make the results highly sensitive to the specific AGN types included in the training set. Following \citet{decicco21,decicco25}, we therefore adopted the leave-one-out cross-validation (LOOCV; \citealt{loocv}), which treats each source in the LS as a single-instance validation set, while the remainder is used for training. This process yields a prediction for every object in the LS, effectively using the full sample for both training and validation without overlap.

The first step in our selection process is to select all the sources classified as AGN in the LS, for each of the tested models, and evaluate the performance of the classifier. We report some classic evaluation metrics for these models in sections I and II of Table \ref{tab:confusion_matrices}. The table includes two of the usual classes for a confusion matrix, namely, true positive ratio (TPR), true negative ratio (TNR)\footnote{True positives (TPs): known AGN correctly classified; true negatives (TNs): known non-AGN correctly classified. A confusion matrix also reports the number of false positives (FPs): known non-AGN incorrectly classified, and the number of false negatives (FNs): known AGN incorrectly classified. For each quantity we can obtain the corresponding ratio with respect to the total number of AGN or non-AGN.}; it also reports accuracy, precision, F1, and recall\footnote{Accuracy $A = \frac{\mbox{TPs}+\mbox{TNs}}{\mbox{Tot. Sample}}$; precision $P = \frac{\mbox{TPs}}{\mbox{TPs}+\mbox{FPs}}$, $F1 = 2\times\frac{P\times R}{P+R}$, recall $R = \frac{\mbox{TPs}}{\mbox{TPs}+\mbox{FNs}}$.}, the last one also being computed for various sub-classes of AGN, defined on the basis of their selection method. Section~I of the table refers to the three tests on the spec-MIR LS, which does not include AGN confirmed only via their X-ray properties, while the main LS, used in the three tests in Section II of the table, includes these sources. Each percentage reported in the table is obtained as the average of ten experiments.

The tests on the spec-MIR LS do not show any significant differences among the three bands, except for a a higher value for the overall recall in the $g$ band, as well as for its various sub-classes. This might be ascribed to the faster variability of AGN in the $g$ band.

The tests on the main LS mostly show the same trend for all the metrics but the recall, the $r$ band mostly returning slightly better performance. For what concerns recall, we note that the overall and most of the sub-class values are consistent within their uncertainties, while we obtained slightly higher values in the $g$ band for obscured AGN and in the $i$ band for MIR AGN. We also note that the X-ray AGN recall is basically independent on the band.

If we compare the two upper sections of the table in order to assess the effect of the inclusion of the X-ray only AGN in the LS, we notice that, essentially, they increase the heterogeneity of the LS, which worsens the TPR as well as the TNR. This leads, as expected, to a general worsening of all the metrics. While the recall of unobscured AGN and MIR AGN does not change dramatically, we do observe a decreased X-ray recall. We stress that, when using the spec-MIR LS, the sub-sample of AGN that are confirmed via X-ray properties is dominated by spectroscopically confirmed AGN, as they make up 305 out of the 315 sources in the X-ray LS, while the remaining 10 are AGN confirmed via their MIR properties. When we use the main LS and hence include X-ray-only AGN, on the other hand, we include potentially different sources, their only known property as AGN being their $X/O$; as a consequence, we interpret this decreased X-ray recall as another effect of heterogeneity. 
The most striking consequence of expanding the LS is, nevertheless, the increase in the recall of obscured AGN, which is around $50\%$ when using the spec-MIR LS and rises to $> 70\%$ values when resorting to the main LS. In spite of the generally lower performance of the classifiers based on the main LS, we stress that a diversified LS represents the most interesting case scenario, as it is what we commonly have to deal with if we aim at including as many source types as possible in our analysis. Hence, the main LS is the one we will focus our further work on in the next sections.

\begin{table*}[tb]
\centering
\caption{\footnotesize{Confusion matrix values for the various RF classifiers tested in the $gri$ bands.}} 
\label{tab:confusion_matrices}  
\renewcommand{\arraystretch}{1.5}
    \resizebox{.99\textwidth}{!}{
\begin{tabular}{c l l c c c c c c c c c c}
\toprule
\ & Band & LS & features & TPR & TNR & accuracy & precision & F1 & unobscured AGN & obscured AGN & X-ray & MIR\\
\ & & & & (recall)                               &     &     &           & & recall  & recall & recall & recall\\
\midrule
\  &$g$ & spec-MIR & all & $81.6\pm0.2$ & $98.18\pm0.16$ & $95.42\pm0.14$ & $89.9\pm0.8$ & $85.6\pm0.4$ & $99.7\pm0.2$ & $53.2\pm0.6$ & $85.7\pm0.3$ & $88.2\pm0.4$\\
\ \textbf{I} & $r$ & spec-MIR  & all & $79.3\pm0.8$ & $98.74\pm0.15$ & $95.52\pm0.15$ & $92.6\pm0.8$ & $85.5\pm0.5$ & $99.00\pm0.15$ & $50.1\pm1.5$ & $84.2\pm0.5$ & $86.4\pm1.0$\\
\ & $i$ & spec-MIR & all & $79.2\pm0.6$ & $98.03\pm0.10$ & $94.90\pm0.16$ & $88.9\pm0.5$ & $83.8\pm0.5$ & $99.2\pm0.2$ & $46.4\pm1.5$ & $83.2\pm0.6$ & $87.9\pm0.7$\\
\midrule
\ & $g$ & main &  all &$74.8\pm0.4$ & $95.72\pm0.15$ & $91.03\pm0.11$ & $83.5\pm0.4$ & $78.7\pm0.2$ & $99.5\pm0.0$ & $73.5\pm1.8$ & $76.3\pm0.4$ & $89.2\pm0.3$\\
\ \textbf{II} &$r$ & main & all &$74.3\pm0.5$ & $96.4\pm0.2$ & $91.4\pm0.2$ & $85.7\pm0.7$ & $79.4\pm0.5$ & $99.6\pm0.2$ & $71\pm2$ & $76.2\pm0.5$ & $89.5\pm0.6$\\
\ & $i$ & main &  all & $74.9\pm0.9$ & $95.0\pm0.2$ & $90.5\pm0.2$ & $81.4\pm0.6$ & $77.8\pm0.6$ & $99.6\pm0.2$ & $72.8\pm1.9$ & $76.3\pm0.9$ & $91.5\pm0.6$\\
\midrule
\ & $g$ & main & var. features only & $47.69\pm0.18$ & $98.28\pm0.14$ & $86.92\pm0.13$ & $88.9\pm0.8$ & $62.1\pm0.3$ & $92.30\pm0.15$ & $29.2\pm0.0$ & $51.6\pm0.2$ & $66.45\pm0.16$\\
\ \textbf{III} & $r$ & main & var. features only & $51.6\pm0.2$ & $99.31\pm0.07$ & $88.60\pm0.07$ & $95.6\pm0.4$ & $67.0\pm0.2$ & $97.0\pm0.2$ & $37.4\pm0.3$ & $55.6\pm0.2$ & $70.9\pm0.0$\\
\ & $i$ & main & var. features only & $47.0\pm0.5$ & $95.5\pm0.3$ & $84.6\pm0.3$ & $75.3\pm1.3$ & $57.8\pm0.6$ & $88.3\pm0.7$ & $24.5\pm1.1$ & $49.5\pm0.5$ & $69.5\pm1.0$\\
\midrule
\ \textbf{IV} & - & main & \emph{uBrizy}-derived colors only & $73.8\pm0.3$ & $89.1\pm0.3$ & $85.7\pm0.2$ & $66.3\pm0.6$ & $69.9\pm0.4$ & $96.0\pm0.3$ & $64.6\pm1.0$ & $75.9\pm0.4$ & $82.7\pm0.6$\\
\ & - & main & all colors only & $76.0\pm0.8$ & $93.8\pm0.2$ & $89.8\pm0.3$ & $78.1\pm0.7$ & $77.0\pm0.7$ & $99.3\pm0.3$ & $68\pm3$ & $76.6\pm0.9$ & $93.3\pm0.5$\\
\midrule
\ & $g$ & main &  13, optimized for obscured AGN sel. & $77.8\pm0.8$ & $94.76\pm0.14$ & $90.96\pm0.15$ & $81.1\pm0.4$ & $79.5\pm0.4$ & $99.95\pm0.15$ & $81\pm2$ & $79.2\pm0.7$ & $91.8\pm0.6$\\
\ \textbf{V} & $r$ & main & 14, optimized for obscured AGN sel. & $76.7\pm0.6$ & $95.7\pm0.2$ & $91.4\pm0.2$ & $83.7\pm0.7$ & $80.0\pm0.5$ & $99.5\pm0.0$ & $79.0\pm1.8$ & $78.\pm0.7$ & $90.0\pm0.3$\\
\ & $i$ & main & 10, optimized for obscured AGN sel. & $80.1\pm0.2$ & $94.72\pm0.17$ & $91.42\pm0.14$ & $81.4\pm0.5$ & $80.7\pm0.3$ & $99.5\pm0.0$ & $84.3\pm0.9$ & $81.1\pm0.2$ & $93.3\pm0.2$\\
\midrule
\bottomrule
\end{tabular}
}
\\
\vspace{0.5em}
\footnotesize{\textbf{Notes.} The table reports information about the band, LS, and features used for each test, as well as true positive ratio (TPR), true negative ratio (TNR), accuracy, overall precision values, F1, and recall for unobscured and obscured AGN, X-ray AGN, and MIR AGN. The overall value of the recall is by definition the same as the TPR. Sections I and II report the results obtained using the two different LSs adopted in this work. Sections III and IV report the results from the tests where only variability features or only colors are used, respectively, which will be discussed in Section \ref{section:colors}. Section V is focused on the selection of obscured AGN (se also Table \ref{tab:agn2_feats}), which will be discussed in Section \ref{section:agn2}. All values are to be read as percent values. The percentage errors represent the standard deviation from the mean value derived from a set of ten experiments per classifier. The Roman numerals in the leftmost column help identify the various sections of the table throughout this work.}
\end{table*}

\section{Models using optical variability or colors alone}
\label{section:colors}
As also mentioned in the introduction to this work, it is well known that in obscured AGN the nuclear continuum and broad-line region are hidden by circumnuclear material, so that the observed optical emission is dominated by the host galaxy. As a result, the intrinsic variability signal is heavily diluted or completely suppressed, making obscured AGN much harder to recover through optical variability-based selection alone. In previous works in this series \citep{decicco21,decicco25} we discussed the primary role of color features for improving the identification of obscured AGN. This naturally raises the question of whether it would be possible to rely solely on color features for classification, thereby avoiding the complexity of time-domain data, light curves, and all the associated challenges. In what follows, we report on two complementary tests in which we constructed different RF classifiers using, respectively, variability features alone in each available band, or color features alone. The comparison with the previous tests in this work assesses the distinct advantages and shortcomings of these approaches in the context of AGN selection.

The first set of experiments in this section uses only the 29 variability features reported in Table \ref{tab:features}. The RF classifier is trained on the main LS. Once again, we followed our standard hyperparameter optimization procedure. All the experiments limited to variability features only yielded remarkably worse results. Section III of Table \ref{tab:confusion_matrices} presents the metrics obtained when using the full set of 29 variability features from Table \ref{tab:features}. Compared to Section II, where the same LS and the full set of features were used, a slight enhancement is observed only in the TNR, but this comes at the cost of  marked declines in the TPR, F1 score, and all the recall values. In particular, the recall for obscured AGN is notably lower than in any other experiments presented in this work. We also observe that, overall, the recall in the $r$ band is broadly consistent with the findings of \citet{decicco21}. However, we caution that the LS used in that study was less heterogeneous than the one adopted here (as is discussed in Section \ref{section:LS}). A further noteworthy outcome from this analysis is that the $r$ band consistently outperforms the other two bands, which can very likely be attributed to its higher observing cadence. In particular, the $r$-band precision and the TNR from this model are the highest across the whole table, suggesting that the classifier did an excellent job in identifying non-AGN via their optical variability features alone. By contrast, compared to the other two bands, the $i$ band exhibits a larger scatter in its performance, and its metrics are the lowest across all the tests in the table. Nevertheless, this band still confirms useful in identifying MIR AGN.

We then tested a RF classifier trained on the main LS, using only the color features expected to be available from the LSST -- namely, $u$–$B$, $B$–$r$, $r$–$i$, $i$–$z$, and $z$–$y$ -- with no use of variability features. We followed the same hyperparameter optimization procedure adopted throughout this work and combined the results from ten experiments as usual. We reported the outcomes in the first line of Section IV of Table \ref{tab:confusion_matrices}. 
As clearly shown in the table, this purely color-based approach results in an overall significant decline in classifier performance compared to Section II of the table. In particular, we observe that the precision sinks, being $15\%-17\%$ lower, which means the classification will suffer from substantial contamination. This is a typical issue when using only colors, and excluding the variability features typically significantly hampers the classification accuracy and leads to critically contaminated AGN samples \citep[e.g.,][]{decicco21}. We also note an average $6\%$ drop in the TNR across the three bands, which means that, on average, based on the size of our LS we are misclassifying $>100$ more non-AGN compared to previous experiments. The recall for obscured AGN is also lower by $\approx8\%$ on average across the three bands.

We also tested the use of all six color features from Table~\ref{tab:features}, adding the MIR color feature \texttt{ch21} to the previous experiments since, as was mentioned above, this is well known to be crucial for the identification of obscured AGN. The obtained results are reported in the second line of Section IV of Table \ref{tab:confusion_matrices}. In this case the results are much better than in the previous line in the table but, overall, the exclusion of the variability features worsens the performance compared to Section II (especially the $g$ and $r$- band-related lines), pointing toward a higher contamination. Only the TPR and, as expected, the recall for MIR AGN are higher than in Section II.

The results analyzed in this section aim at confirming that neither optical variability features alone nor color features alone are as effective in selecting AGN as their combination is; and, in particular, if we focus on obscured AGN we find that so far the highest recall has been returned by the full set of features from Section II of Table \ref{tab:features}. This said, in the next section we will focus on optimizing the selection of obscured AGN.

\section{Selection of obscured AGN}
\label{section:agn2}
In previous works in this series \citep{decicco21, decicco25} we extensively discussed the issues affecting the selection of obscured AGN via optical variability and their generally much lower completeness compared to unobscured AGN. \citet{decicco25} also showed how to identify an optimized set of features to select obscured AGN and obtained a more complete sample than previous works in the series, though still affected by a large contamination. Specifically, that previous analysis explored whether combining the $g$- and $r$-band features could improve classifier performance. In order to do so, the analysis was limited to 33 visits common to the two bands, where some gaps in either band had been filled via data imputation.
The approach, however, did not result in any significant gain for the selection of obscured AGN, as the best-performing feature subsample ultimately included, among others, variability features from one band only, this being the $r$. Here we therefore adopted a different strategy, using the data from the three available bands -- $g$, $r$, and $i$ -- independently to select AGN in each of them. This means that here we use all the available visits, without introducing artificial data to account for missing visits. This approach will be beneficial for LSST data where different bands are likely to be sampled with different cadences, although these will be higher than VST cadences. We finally merged these independent results in order to obtain a cleaner and hence more robust sample.

We have seen that our reference sample of obscured AGN consists of 96 sources detected in the X-rays and classified as Type II via optical spectroscopy. We point out that these sources were also used in \citet{decicco22}, where the structure function of various subsamples of AGN selected via different properties and methods was analyzed. In that case, the sample of Type II AGN consisted of 104 sources, and thus included eight additional sources that we had to exclude from the present work based on the adopted cuts, as is detailed in Section \ref{section:LS}. Nevertheless, 92\% of the sample of Type II AGN used in \citet{decicco22} consisted of the same Type II AGN here used, and that work clearly shows (e.g., in their Fig. 5) that this subsample of sources appears to be less variable than all the other subsamples of AGN there analyzed. This suggests that, while the Type~II classification of these sources may be debated based on the low signal-to-noise ratio of the original spectra and the fact that they were observed at a different time with respect to the VST light curves used here, they are still a peculiar subsample of objects for what concerns their optical variability, less variable and harder to identify, and are therefore worth being investigated separately.

As we mentioned earlier, each of the values reported in Table~\ref{tab:confusion_matrices} is the average outcome of a series of ten experiments. As a consequence, each of these experiments will provide an independent classification for each source in the LS. If we focus on obscured AGN, we can combine the results from the ten experiments corresponding to each tested model, and obtain a final classification as AGN or non-AGN for each source. In order to do so, we need to set a classification threshold. In the top and middle sections of Table \ref{tab:intersection} we report the results obtained on the basis of three different thresholds, for the first two series of ten experiments testing two different LSs: first, we considered a source to be an AGN only if all the ten experiments corresponding to a given model classify it as an AGN (10/10 threshold); then we released this criterion and lowered the threshold to eight out of ten positive outcomes (8/10 threshold), and finally we tested an even less conservative approach, adopting a 6/10 threshold. Once we obtained the final classification for each source per band, we cross-matched the results and obtained a final sample of sources classified as AGN in all three bands, for each of the tested thresholds. {We also reported the number of sources classified as AGN in two out of the three bands, as well as} the union of the results, i.e., the number of sources classified as AGN in at least one band. This last number is to be considered as an upper limit to the completeness that we can achieve based on the used data and feature sets and we stress that, although some sources are classified as AGN only in one band, this classification is still the result of multiple (10, 8 or 6) tests.

From the table we can see that, for the tests based on the spec-MIR LS, the final results intersecting the samples obtained for each band are not very different from each other, regardless the adopted threshold; on the contrary, the threshold seems to matter more in tests using the main LS. Based on these results, we decided to build our further analysis on the test performed using the main LS, and with 8/10 as a threshold; indeed, this LS returns a much improved recall for obscured AGN, and the threshold is a representative average benchmark for the following discussion, since especially for the union the results do not change much.

\begin{table}[tb]
\centering
\caption{\footnotesize{Number of obscured AGN confirmed from each tested model, in each band.}} 
\label{tab:intersection}  
\renewcommand{\arraystretch}{1.5}
     \resizebox{.99\columnwidth}{!}{
\begin{tabular}{l c c c c c c}
\toprule
\ spec-MIR LS & $g$ & $r$ & $i$ & and ($\cap$) & 2 out of 3 bands & or ($\cup$)\\
\ 10/10 & 49 & 44 & 40 & 33 ($34^{+7}_{-6}\%$) & 46 ($48^{+8}_{-7}\%$) & 55 ($57^{+10}_{-9}\%$)\\
\ 8/10 & 51 & 46 & 43 & 36 ($38^{+7}_{-6}\%$) & 48 ($50^{+8}_{-8}\%$) & 57 ($59^{+9}_{-8}\%$)\\
\ 6/10 & 51 & 47 & 44 & 36 ($38^{+7}_{-6}\%$) & 49 ($51^{+8}_{-7}\%$) & 59 ($61^{+9}_{-8}\%$)\\
\midrule
main LS & & & & & \\
\ 10/10 & 65 & 62 & 59 & 49 ($51^{+8}_{-7}\%$) & 61 ($64^{+9}_{-8}\%$) & 80 ($83^{+10}_{-9}\%$)\\
\ 8/10 & 68 & 65 & 65 & 56 ($58^{+9}_{-8}\%$) & 66 ($69^{+10}_{-8}\%$) & 79 ($82^{+10}_{-9}\%$)\\
\ 6/10 & 71 & 67 & 68 & 60 ($62^{+9}_{-8}\%$) & 68 ($71^{+10}_{-9}\%$) & 79 ($82^{+10}_{-9}\%$)\\
\midrule
main LS, & & & &  & \\
\ obscured AGN-focused & & & &  & \\
\ 10/10 & 69 & 69 & 78 & 60 ($62^{+9}_{-8}\%$) & 74 ($77^{+10}_{-9}\%$) & 89 ($93^{+11}_{-10}\%$)\\
\ 8/10 & 74 & 72 & 81 & 66 ($69^{+10}_{-8}\%$) & 77 ($80^{+10}_{-9}\%$) & 88 ($92^{+11}_{-10}\%$)\\
\ 6/10 & 79 & 76 & 81 & 72 ($75^{+10}_{-9}\%$) & 79 ($82^{+10}_{-9}\%$) & 87 ($91^{+11}_{-10}\%$)\\
\bottomrule
\end{tabular}
 }\\
\vspace{0.5em}
\footnotesize{Notes. The top section refers to tests using the spec-MIR LS, i.e., not including AGN confirmed only via their X-ray properties. The middle section refers to tests using the main LS, which includes these AGN. The bottom section refers to additional tests discussed later in this section, where we used the main LS and a specific selection of features for each band, aimed at optimizing the selection of obscured AGN. The various columns indicate: the adopted threshold (1); the three bands (2-4), where the final number for each model tested was obtained combining the outcomes of the ten independent experiments, according to the threshold indicated by each line; the intersection of the results (5), i.e., the number (and percentage) of obscured AGN confirmed in all three bands; the number (and percentage) of obscured AGN confirmed in two out of three bands; and the union of the results (6), i.e., the number (and percentage) of sources classified as obscured AGN in at least one band. Error bars on the reported percentages were estimated following the approach of \citet{roman}, who provided approximated formulae for $95\%$ confidence intervals under the assumptions of Poisson and binomial statistics. Percentages were computed with respect to the total number of obscured AGN, these being 96.}
\end{table}

\begin{figure}[t]
\centering
    {\includegraphics[width=9cm]{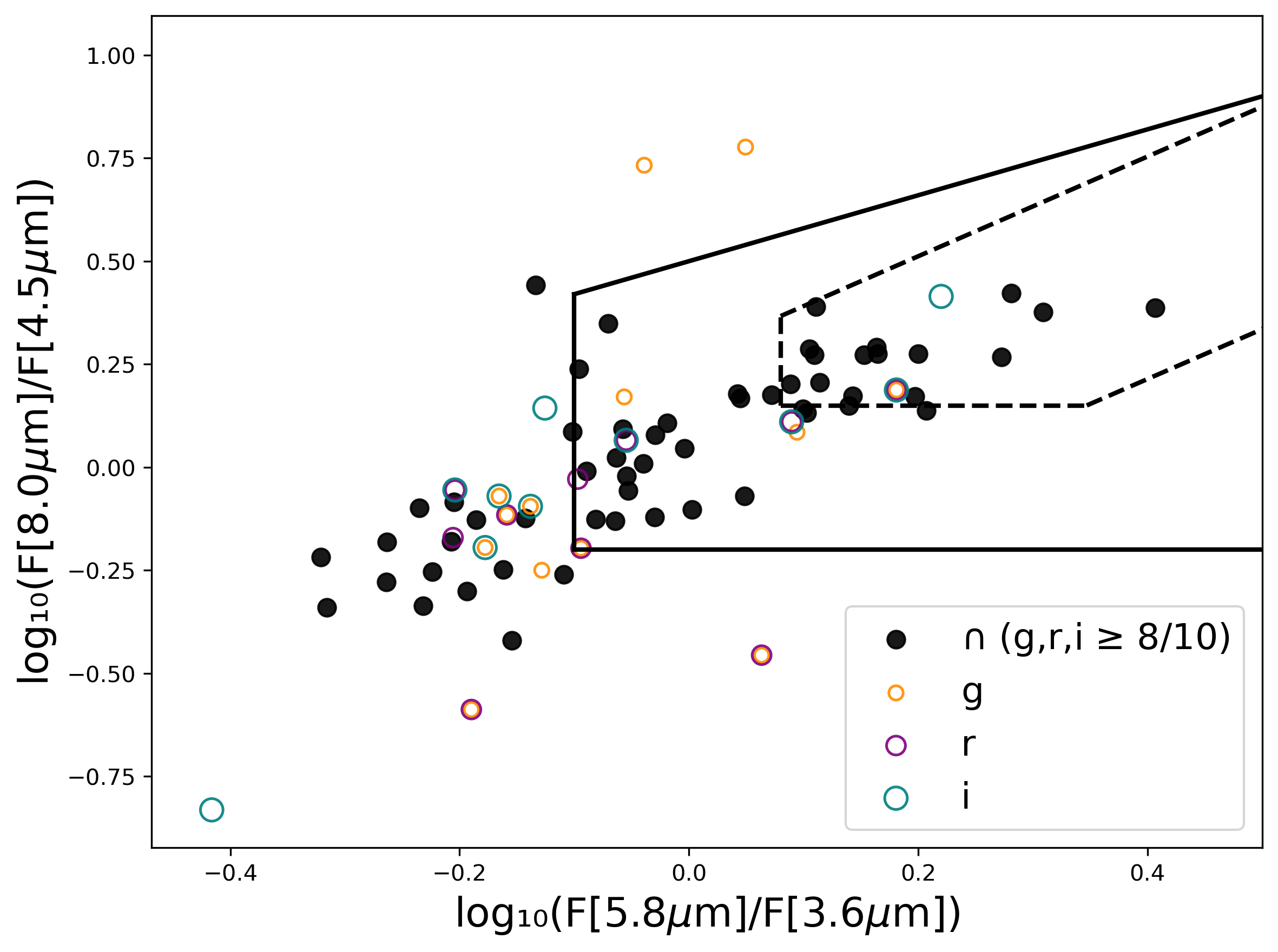}}
\caption{\footnotesize Mid-infrared color-color diagram for our sample of 96 obscured AGN. The solid and dashed lines define the region where AGN are typically found after \citet{lacy07} and \citet{donley}, respectively. Filled black dots indicate sources that were classified as AGN in all three bands with an 8/10 threshold, while empty circles indicate sources classified as AGN in one or two bands only (using the same 8/10 threshold in each band), according to the legend in the figure.}\label{fig:MIR}
\end{figure}

Since MIR information is available from the COSMOS2015 catalog \citep{laigle} for the whole sample of 96 obscured AGN used in this work, we reported them on a diagram in Fig.~\ref{fig:MIR}, comparing two MIR colors obtained from the four IRAC filters corresponding to 3.6 $\mu m$, 4.5 $\mu m$, 5.8 $\mu m$, and 8.0 $\mu m$. The figure includes lines delimiting typical AGN loci based on \citet{lacy07} and \citet{donley}. It is apparent that a significant fraction of the obscured AGN in our sample do not lie in these regions, which is both intriguing and challenging as it shows how different AGN selection techniques are complementary: indeed, in this case it shows how selection based mostly on optical variability properties combined with colors is able to unearth AGN that would be otherwise missed via MIR-based selection, as they lie in the area where star-forming galaxies are typically found. Indeed, \citet{donley} discuss how in this region we can find obscured AGN that will be mostly missed via X-ray or MIR selection. Another point is that, since the MIR data we used are not contemporary to our VST data, the state of at least some of these sources might have changed; if so, they would show different properties in optical and MIR wavelengths.

One of the primary challenges in identifying obscured AGN lies in distinguishing them from inactive galaxies. Following the approach outlined in \citet{decicco25}, we analyzed the feature sets in each band to assess which ones are most effective in separating these two classes of sources. For each feature, we therefore compared the distributions corresponding to obscured AGN and inactive galaxies, applying a Kolmogorov-Smirnov (K-S) test to identify the features whose distributions differ significantly and for which the probability of obtaining by chance a larger difference is low. Based on the K-S test results in each band, we defined the corresponding thresholds and thus identified a subset of discriminative features to be used in the subsequent analysis. 
We stress that, while the procedure closely follows the approach adopted in \citet{decicco25}, there are some key differences here: first of all, since here we are using the three bands independently, the light curve-derived features for each band are used separately. This allows us to make full use of the entire dataset, exploiting all the available visits, without the need for any artificial adjustments. Conversely, \citet{decicco25} used $r$, $g$, and also bivariate features together, which made it necessary to resort to data imputation for the light curves in the two bands so that all the features used together had the same weight. Another difference is that here we focus on an AGN LS that is $\approx1.4$ times larger than the previous one (520 versus 380 AGN) and also more diversified, better reflecting a realistic scenario.

Essentially, for each band we used the selected subset of features and repeated the hyperparameter optimization procedure described in Section \ref{section:classification}. We then trained a classifier using only these features and performed ten independent experiments, as usual. It is important to note that each classifier was trained on the whole main LS and not only on obscured AGN and inactive galaxies. From each set of ten experiments, we derived a feature importance ranking and subsequently removed the least informative feature, repeating the entire process iteratively. Feature importance was evaluated as the average impurity decrease across all the trees where the feature was used and involved in a split. We assessed the performance of each classifier using the same metrics reported in Table \ref{tab:confusion_matrices}, and halted the iteration when a further reduction in the number of features no longer yielded any performance gains. This procedure, applied independently to the $g$, $r$, and $i$ bands, revealed that the best-performing classifiers rely on 13, 14, and 10 features in the $g$, $r$, and $i$ bands, respectively. We listed the selected features for each band in Table \ref{tab:agn2_feats}, while we reported the corresponding classifier performance metrics in Section V of Table \ref{tab:confusion_matrices}. 

Table \ref{tab:agn2_feats} shows that eight features consistently appear in the selected lists across the three bands: these are the variability features \texttt{GP\_DRW\_$\tau$} and \texttt{$\eta^e$},
the colors \texttt{ch21}, \texttt{u-B}, \texttt{r-i}, \texttt{i-z}, \texttt{z-y} (i.e., all the color features considered except for \texttt{B–r}), and the morphology indicator \texttt{class\_star}. Moreover, four additional variability features -- \texttt{ExcessVar}, \texttt{IAR$_\phi$}, \texttt{Autocor\_length}, \texttt{R$_{cs}$} -- are present in the selected feature sets for two out of the three bands. These results suggest that, although the exact feature set always depends on the specific band and on the LS used, certain features consistently play a central role in the classification process. Interestingly, we note that several variability features reported in Table \ref{tab:agn2_feats} (e.g., \texttt{GP\_DRW\_$\tau$}, \texttt{IAR$_\phi$}, \texttt{Autocor\_length},  \texttt{R$_{cs}$}) are related to the timescale of the variations and are sensitive to the number of visits, and hence to the light curve sampling. 

In \citet{decicco25} we applied the same optimization procedure starting from 25 features (see their Fig. 3) and identified a final subset of eight features (see their Fig. 4) that proved most effective in separating obscured AGN and inactive galaxies. As was mentioned above, although the initial set of 25 features included both $r$- and $g$-band features, the final subset of eight contained only $r$-band features, together with five colors and the morphology indicator. 
For comparison, all the 14 features selected as most effective in the present work are included in the initial selection of 25 features used in \citet{decicco25}, and all eight of their final features are also found among our final subset of 14, with half of them also occupying the same rank position. Despite such a consistency, the fact that we had started from a different feature sample and had trained the algorithm on a different LS naturally led to a different set of features in the end.

Section V of Table \ref{tab:confusion_matrices} indicates that, in most cases, the results obtained across the three bands via the corresponding subsamples of features are mutually consistent. We note that the performance of the $i$ band is generally slightly better than the other two, and this is interesting especially if we focus on the results obtained for the subsamples of obscured AGN and MIR AGN (where the latter potentially include a significant fraction of obscured AGN): indeed, this supports the well-established notion that redder bands are more effective in the identification of obscured sources. However, the recall values across the various subclasses of sources do not differ dramatically from each other.

A comparison with Section II of the table -- where the same LS was used, but with the entire initial set of features -- reveals that reducing the number of features generally leads to an improved performance in each band. Nonetheless, we consistently notice slightly lower values for TNR and precision across all three bands. We stress that, given the size and composition of our LS, the average $<1\%$ decrease in TNR observed translates into $<20$ additional misclassified sources. An increase in contamination is typically the cost of achieving a higher completeness, which is indeed reflected in our results, as apparent from the various recall values in the table. In particular, in this case the $<1\%$ higher contamination is the price to pay to obtain a  recall for obscured AGN increased by $\approx8\%$ in all three bands. We note that the most complete and most contaminated performance is associated with the $i$-band classifier. This could, at least in part, result from the less efficient cadence in this band, which is likely to weaken the variability features \citep[see, e.g., Fig. 10 of][]{decicco19}. We therefore expect this contamination to be reduced when using LSST data. We stress how, in this set of experiments focused on obscured AGN, a more diverse and realistic LS led to a generally worse performance but substantially improved the recall of obscured AGN.

In line with the approach adopted earlier in this work, we set a probability threshold of eight out of ten experiments as our reference, which yields single-band samples of obscured AGN larger than any other experiment here performed so far. Specifically, we were able to retrieve 74 ($g$ band), 72 ($r$ band), and 81 ($i$ band) out of the 96 known obscured AGN in our LS, which translate into $77^{+10}_{-9}\%$, $75^{+10}_{-9}\%$, and $84^{+10}_{-9}\%$, respectively. 
We also combined the results of the ten experiments performed for each band with the corresponding selected set of features. These results are presented in the bottom section of Table \ref{tab:intersection}, including the intersection and union of the retrieved samples as well as the number of sources confirmed in two out of three bands. We can see that a sample of 66 out of 96 ($69^{+10}_{-8}\%$) obscured AGN are correctly classified in all three bands, while 77 AGN ($80^{+9}_{-7}\%$) are correctly classified in two out of three bands.

\begin{table}[tb]
\caption{\footnotesize{List of features yielding the best-performing classifier in each band.}} 
\label{tab:agn2_feats}  
\renewcommand{\arraystretch}{1.5}
\centering

    \resizebox{.8\columnwidth}{!}{
\begin{tabular}{l c c c}
\toprule
\ & $g$ & $r$ & $i$ \\
\ 1 & \texttt{ch21} & \texttt{ch21} & \texttt{ch21}\\          
\ 2 & \texttt{u-B} & \texttt{u-B} & \texttt{u-B}\\       
\ 3 & \texttt{class\_star} & \texttt{class\_star} & \texttt{class\_star}\\
\ 4 & \texttt{r-i} & \texttt{GP\_DRW\_$\tau$} & \texttt{B-r}\\ 
\ 5 & \texttt{GP\_DRW\_$\tau$} & \texttt{r-i} & \texttt{r-i}\\   	
\ 6 & \texttt{$\eta^e$} & \texttt{ExcessVar} & \texttt{i-z}\\          
\ 7 & \texttt{IAR$_\phi$} & \texttt{P$_{var}$}& \texttt{GP\_DRW\_$\tau$}\\       
\ 8 & \texttt{i-z} & \texttt{i-z} & \texttt{IAR$_\phi$}\\           
\ 9 & \texttt{R$_{cs}$} & \texttt{R$_{cs}$} & \texttt{z-y}\\           
\ 10 & \texttt{z-y} & \texttt{GP\_DRW\_$\sigma$} & \texttt{$\eta^e$}\\            
\ 11 & \texttt{ExcessVar} & \texttt{z-y} & -\\   
\ 12 & \texttt{Meanvariance} & \texttt{Period\_fit} & -\\   
\ 13 & \texttt{Autocor\_length} & \texttt{$\eta^e$} & -\\
\ 14 & - & \texttt{Autocor\_length} & -\\

\bottomrule
\end{tabular}
}
\\
\footnotesize{\textbf{Notes.} The classifiers are evaluated using the standard performance metrics adopted in this work, with particular emphasis on the selection of obscured AGN. Within each band, features are ranked in order of decreasing importance.}
\end{table}

\begin{figure*}[t]
\centering
\subfigure
            {\includegraphics[width=\columnwidth]{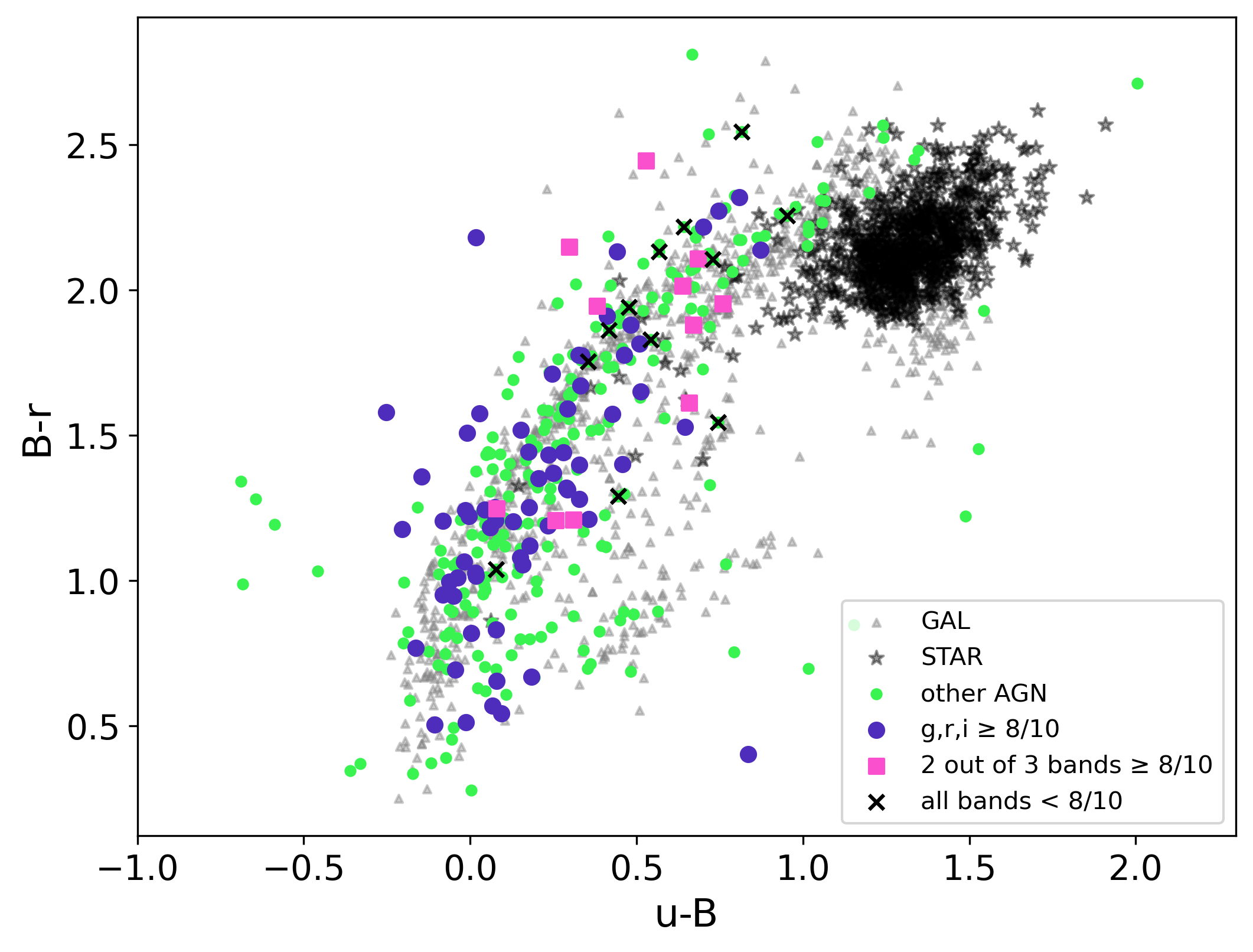}}
\subfigure
            {\includegraphics[width=\columnwidth]{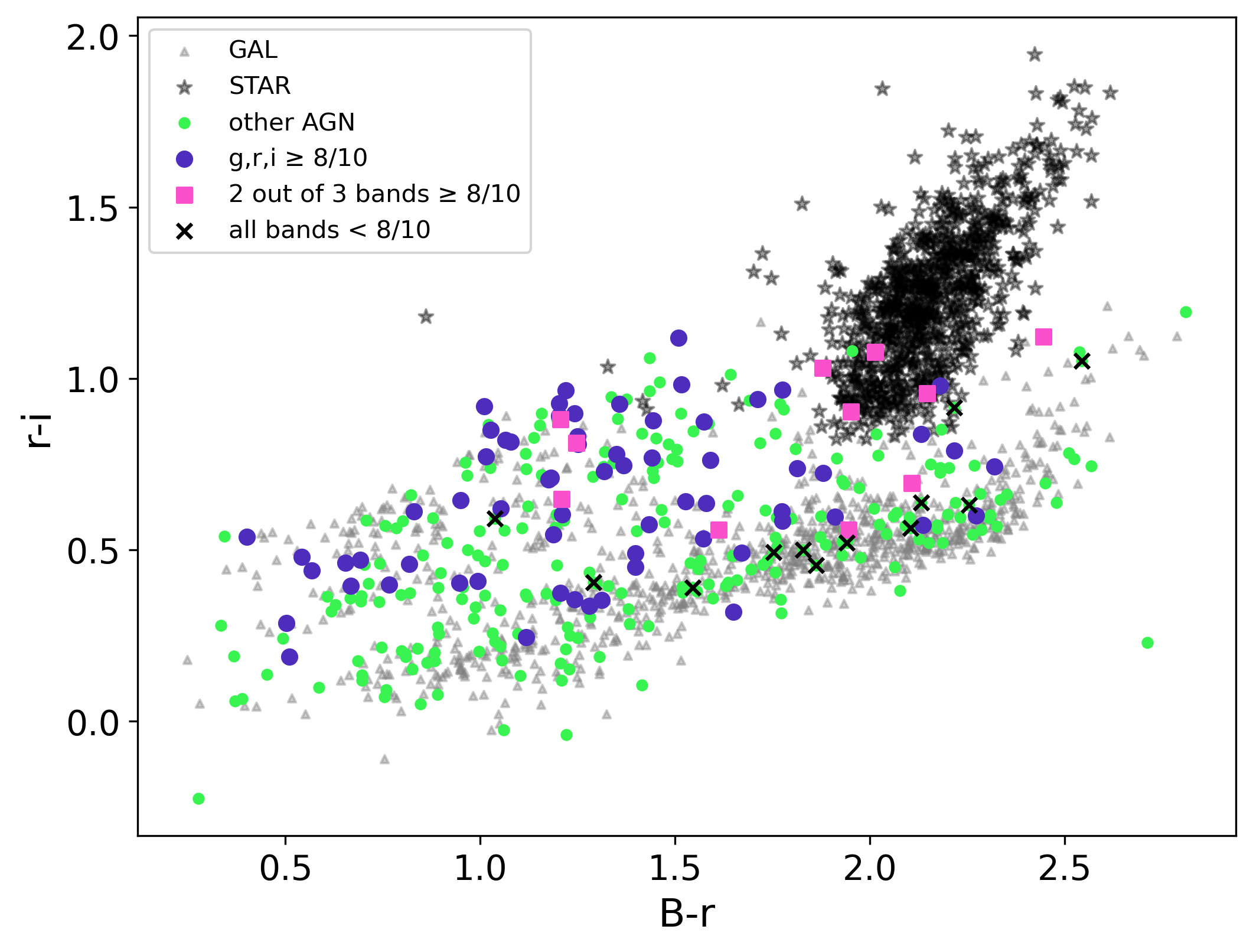}}
\caption{\footnotesize Color-color diagrams where the obscured AGN that were not identified in any of the three bands (black crosses) tend to place themselves on specific loci. In both panels, the gray triangles represent inactive galaxies from the LS, while black stars represent the stars in the LS. The large violet dots indicate the obscured AGN confirmed in all three bands by at least eight out of ten experiments, while the bright pink squares indicate the ones confirmed in two out of the three bands based on the same threshold. The small green dots stand for all the AGN in the LS that are not labeled as obscured.}\label{fig:color_diagrams}

\end{figure*}

In Fig. \ref{fig:color_diagrams} we present two color-color diagrams, namely $B-r$ versus $u-B$ and $r-i$ versus $B-r$, where the sources of the main LS are shown. The left diagram suggests that the obscured AGN that we consistently failed to identify in the three bands are redder than most of the other obscured AGN in the LS, while the right diagram shows that these sources occupy the lower galaxy sequence, having redder $B-r$ but bluer $r-i$ colors. Nevertheless, based on these features alone they do not seem to constitute a separate population with respect to the ``other AGN'' subsample in our LS, that is to say all the AGN in the LS that are not labeled as obscured; hence these diagrams alone are not sufficient to disentangle them. We did not include other color-color diagrams based on other filters as they do not add relevant information.

\begin{figure}[t]
\centering
	\includegraphics[width=\columnwidth]{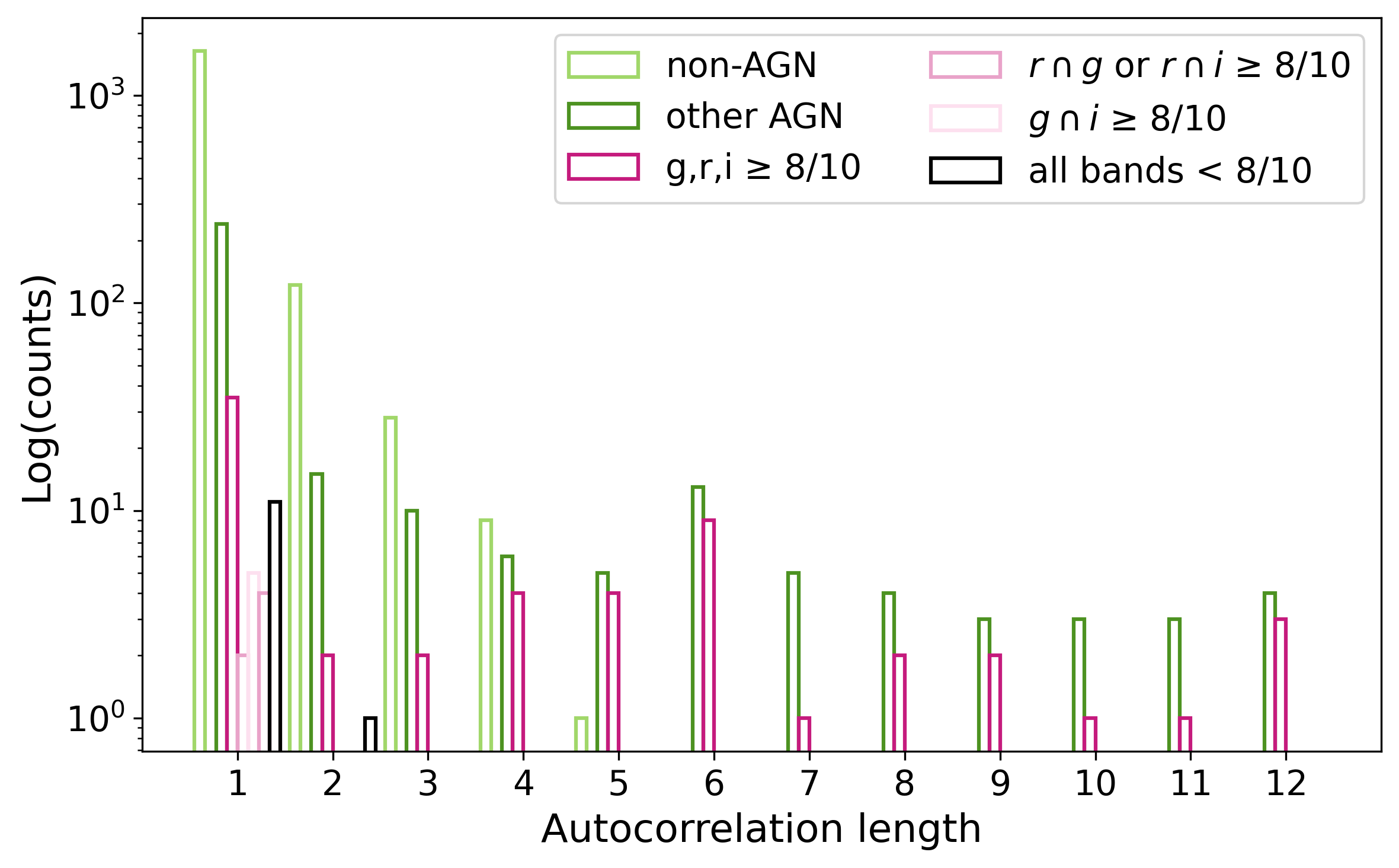}
\caption{\footnotesize Distribution of the values  obtained from the $r$-band light curves for the \texttt{Autocor\_length} feature. The AGN in the LS are split into obscured AGN confirmed in all three bands (deep magenta), in the $r$ band plus either $g$ or $i$ (light magenta), or in the $g$ and $i$ bands, but not in the $r$ (light pink). Dark green indicates all the other AGN, meaning all the AGN in the LS that are not labeled as obscured, while the non-AGN are shown in light green. Black represents the obscured AGN that each classifier consistently failed to identify (that is to say, they were classified as AGN in $<8/10$ experiments per band).}\label{fig:autocorr_l}

\end{figure}

Similarly to colors, the variability features alone generally fail to disentangle the obscured AGN that we are not able to confirm in any of the three bands from inactive galaxies. This can be clearly seen in Fig. \ref{fig:autocorr_l}, where we show the distribution of the feature \texttt{Autocor\_length}, computed from the $r$-band light curves, for the AGN and non-AGN in our LS, distinguishing the various classes of sources. It is indeed apparent that the distribution for obscured AGN identified in all three bands largely overlaps the distribution for all the other AGN, reaching higher values for the feature at issue compared to non-AGN. Things change when a source is not classified as an AGN in the $r$ band or is an obscured AGN that our classifier failed to identify in any of the bands, as the corresponding \texttt{Autocor\_length} values are the lowest possible. While this feature is the least important in the ranking obtained when optimizing selection in the $r$ band (see Table  \ref{tab:agn2_feats}), it provides a very effective visualization of the limitations of optical variability in finding some obscured AGN. 

\section{Summary and conclusions}
\label{section:conclusions}
This work combined the strength of multiband optical variability and color-based selection to identify AGN and, in particular, to mitigate the well-known challenges in detecting obscured sources. The analysis was based on the $g$-, $r$-, and $i$-band light curves extracted from the VST-COSMOS dataset, and focused on the independent identification of AGN in each band. A RF classifier was trained on an initial set of 36 features, of which 29 derived from optical variability, complemented by six optical/infrared colors and a morphology indicator. Since our goal is to provide meaningful insights and practical guidance for future large-scale surveys, the selected features were chosen to reflect those that will be readily accessible for the DDFs from the LSST dataset or from ancillary datasets (for example, the SWIRE/SERVS/DeepDrill surveys will provide magnitudes in the filters necessary to obtain our \texttt{ch21} color feature; see, e.g., \citealt{swire,servs}).

We selected AGN candidates in each band independently, making use of a RF algorithm and testing two different LSs, the difference between the two lying in the AGN populations they include. The so-called spec-MIR LS is composed of spectroscopically confirmed AGN selected from the \emph{Chandra}-COSMOS Legacy Catalog, which is made up of X-ray-detected sources. The objects we selected from this catalog therefore have X-ray information and a spectroscopic classification as Type I or II. The spec-MIR LS also includes MIR-confirmed AGN selected on the basis of the criterion by \citet{donley}. The main LS further includes sources classified as AGN in the \emph{Chandra}-COSMOS Legacy Catalog based solely on their X-ray properties, but with no spectroscopic classification available. Our comparison of the performance of the various classifiers (see sections I and II of Table \ref{tab:confusion_matrices}) shows that the main LS, being more diversified than the spec-MIR LS, leads to an overall decline in performance, with higher contamination rates. Yet, the recall for obscured AGN increases significantly -- by $\approx20\%$ across all bands -- when using the main LS, reflecting the advantage of a more diverse LS in capturing a wider variety of AGN types. Based on the adopted threshold, we are able to identify $58^{+9}_{-8}\%$ of the known obscured AGN. Although this fraction is lower than the $(68.1\pm1.2)\%$ reported in \citet{decicco25}, we note that it is more robust, as the selected sources are consistently classified as AGN in all three bands. We also note (see Fig. \ref{fig:MIR}) that these sources would not be classified as AGN 
based solely on their MIR colors, and this highlights the effectiveness of variability as a complementary selection method to color-based criteria.

We demonstrated that integrating optical variability with color features enhances classifier performance relative to using either set of features separately. Specifically, the classifiers using only optical variability features returned high purity but low completeness, while those using only color features exhibit the opposite trend, achieving higher completeness at the expense of purity.

In the final stage of this study we turned our attention on optimizing the selection of obscured AGN. For each band, we determined the most discriminative features relative to inactive galaxies. We used these feature sets to build a series of classifiers for each band in a recursive way, removing the least informative feature at each step, and stopping when no further gains in the performance were observed. This process yielded a set of 13, 14, and 10 features for the $g$, $r$, and $i$ bands, respectively (see Table~\ref{tab:agn2_feats}). From these results, we derived a sample of AGN candidates in each band, achieving single-band recall values for obscured AGN that exceed those obtained here and in earlier studies in this series.
We subsequently combined the three samples to obtain a more robust final sample. We decided to classify a source as an AGN if it was classified as such in at least eight out of the ten experiments run for each classifier (see Table \ref{tab:intersection}). This strategy resulted in a recall of $69^{+10}_{-8}\%$ for the obscured AGN identified in all three bands. While this is numerically consistent with the findings of \citet{decicco25}, in this work we extended the result to a more comprehensive and diversified sample of AGN classified on the basis of spectra, MIR colors, and X-ray emission. If we adopt a looser criterion, demanding AGN to be confirmed in at least two out of three bands, we obtain an $80^{+10}_{-9}\%$ recall. The obscured AGN samples obtained combining two or three bands should be regarded as purer and more reliable than those derived from single bands. The variability of AGN across different optical bands is known to be correlated, reflecting the physical connection among emission regions located at different distances from the central SMBH. One would therefore expect that a source varying in one optical band also varies in the others, so that optical multiband variability detection alone would not necessarily provide additional information, except reducing contamination as it would remove spurious variability. In our case, however, the classification does not rely on variability alone, but on its combination with color-based properties. Since each classifier uses both variability and color features, their correlations are not trivial. The selection obtained in each band depends on how effectively the variability features computed in that specific band, combined with colors, can identify obscured AGN. As a consequence, detections in more than one band provide complementary information, leading to a more reliable and robust sample of obscured AGN.
Overall, this study proves that a careful selection of features tailored to the identification of obscured AGN, combined with a multiband approach, results in a markedly enhanced recovery rate of known sources. A crucial aspect of this result is that it was achieved by training the classifier via a more diversified LS than in our previous studies. As such, it is more representative of the true AGN population, including sources with a broader range of properties than before.
 
In the future we plan to explore the sources that were excluded from the final sample, in order to verify whether we excluded potentially interesting obscured targets. Indeed, these sources may display variability properties distinct not only from unobscured AGN but also from the selected sample, and these properties could, in principle, vary from band to band. 

Testing this method on the other DDFs, beginning with VST data -- and ultimately extending the analysis to the LSST DDF data, and hence to more bands -- will validate the robustness of the selection technique. An additional key step would be the spectroscopic follow-up of our obscured AGN sample, which could offer valuable insights into both the nature and the physical state of these sources.

\begin{acknowledgements}
DD acknowledges PON R\&I 2021, CUP E65F21002880003, and Fondi di Ricerca di Ateneo (FRA), linea C, progetto TORNADO. DD, MP and VP also acknowledge the financial contribution from PRIN-MIUR 2022 and from the Timedomes grant within the ``INAF 2023 Finanziamento della Ricerca Fondamentale''.
\end{acknowledgements}

\bibliographystyle{aa}
\bibliography{aa56680-25}

\clearpage
\appendix
\onecolumn
%\needspace{10\baselineskip}
\section{Dataset} \label{appendixA}
Here we report detailed information about the three-band datasets used throughout this work. We refer the reader to Section 2 of \citet{decicco15} and Section 2 of \citet{decicco19} for an overview of the data structure and a description of the reduction process.

\begin{table*}[ht]\
\caption{\footnotesize{VST-COSMOS $gri$ dataset.}} 
\label{tab:dataset}      
\hfill

\begin{minipage}{0.497\textwidth}\centering
\begin{tabular}{c c c c c c }
\toprule
\ ID & obs. date & time  & $r$ band & $g$ band & $i$ band\\
\    &           &   (days)          & seeing   & seeing   & seeing  \\   
\midrule
1  & 2011-Dec-18 & 0  & 0.64 & - & -\\
2  & 2011-Dec-22 & 4  & 0.94 & - & -\\
3  & 2011-Dec-27 & 9  & 1.04 & 1.13 & 0.93\\
4  & 2011-Dec-31 & 13 & 1.15 & - & -\\
5  & 2012-Jan-02 & 15 & 0.67 & - & -\\
6  & 2012-Jan-06 & 19 & 0.58 & - & -\\
7  & 2012-Jan-18 & 31 & 0.62 & - & -\\
8  & 2012-Jan-20 & 33 & 0.88 & - & -\\	
9  & 2012-Jan-22 & 35 & 0.81 & 1.11 & 0.95\\	
10 & 2012-Jan-24 & 37 & 0.67 & - & -\\
11 & 2012-Jan-27 & 40 & 0.98 & - & -\\
12 & 2012-Jan-29 & 42 & 0.86 & - & -\\
13 & 2012-Feb-02 & 46 & 0.86 & 0.88 & 1.05\\
14 & 2012-Feb-16 & 60 & 0.50 & 0.61 & 0.60\\
15 & 2012-Feb-19 & 63 & 0.99 & - & -\\
16 & 2012-Feb-21 & 65 & 0.79 & - & -\\
17 & 2012-Feb-23 & 67 & 0.73 & - & -\\
18 & 2012-Feb-26 & 70 & 0.83 & 1.04 & 0.88\\
19 & 2012-Feb-29 & 73 & 0.90 & - & -\\
20 & 2012-Mar-03 & 76 & 0.97 & - & -\\
21 & 2012-Mar-13 & 86 & 0.70 & - & -\\
22 & 2012-Mar-15 & 88 & 1.08 & - & -\\
23 & 2012-Mar-17 & 90 & 0.91 & 1.04 & 0.53\\
24 & 2012-May-08 & 142 & 0.74 & - & -\\	
25 & 2012-May-09 & 143 & -   & 0.78 & 0.62\\
26 & 2012-May-11 & 145 & 0.85 & - & -\\	
27 & 2012-May-17 & 151 & 0.77 & - & -\\	
28 & 2013-Dec-25 & 728 & -   & 0.93 & -\\
29 & 2013-Dec-27 & 740 & 0.72 & - & \textbf{0.70}\\	
30 & 2013-Dec-30 & 743 & 1.00 & - & -\\	
31 & 2014-Jan-03 & 747 & 0.86 & 0.96 & -\\	
32 & 2014-Jan-05 & 749 & 0.81 & - & 0.77\\	
33 & 2014-Jan-12 & 756 & 0.73 & - & 0.73\\	
34 & 2014-Jan-21 & 765 & 1.18 & 1.13 & -\\
35 & 2014-Jan-24 & 768 & 0.80 & - & 0.73\\	
35b & 2014-Feb-02 & 777 & - & - & 0.80\\	
36 & 2014-Feb-09 & 784 & 1.28 & - & 0.98\\
\bottomrule
\end{tabular}
\vspace{5mm}

\end{minipage} 
\vrule width 1pt 
\begin{minipage}{0.497\textwidth}\centering
\begin{tabular}{c c c c c c}
\toprule
\ ID & obs. date & time  & $r$ band & $g$ band & $i$ band\\
\    &           &  (days)           & seeing   & seeing   & seeing  \\   
\midrule
37 & 2014-Feb-19 & 794 & 0.89 & 0.91 & -\\	
38 & 2014-Feb-21 & 796 & 0.93 & - & 0.83\\	
39 & 2014-Feb-23 & 798 & 0.81 & - & -\\	
40 & 2014-Feb-26 & 801 & 0.81 & 0.82 & -\\	
41 & 2014-Feb-28 & 803 & 0.77 & - & 0.91\\	
42 & 2014-Mar-04 & 807 & -    & 1.24 & -\\
43 & 2014-Mar-08 & 811 & 0.91 & - & 0.89\\	
44 & 2014-Mar-21 & 824 & 0.96 & 1.04 & -\\	
45 & 2014-Mar-23 & 826 & 0.92 & - & 0.98\\	
46 & 2014-Mar-25 & 828 & 0.66 & - & -\\
47 & 2014-Mar-29 & 832 & 0.89 & 0.86 & -\\	
48 & 2014-Apr-04 & 838 & \textbf{0.58} & \textbf{0.72} & -\\	
49 & 2014-Apr-07 & 841 & 0.61 & - & -\\	
50 & 2014-Dec-03 & 1081 & 1.00 & - & -\\	
\phantom{a}50b & 2014-Dec-05 & 1083 & - & - & 1.42\\	
\phantom{a}50c & 2014-Dec-15 & 1093 & - & - & 0.69\\	
51 & 2014-Dec-16 & 1094 & -    & 0.80 & -\\
52 & 2014-Dec-25 & 1103 & -    & 0.80 & -\\
53 & 2015-Jan-03 & 1112 & -    & 1.09 & -\\
54 & 2015-Jan-10 & 1119 & 0.71 & - & -\\	
55 & 2015-Jan-15 & 1124 & -    & 0.88 & -\\
\phantom{a}55b & 2015-Jan-21 & 1130 & -    & - & 0.78\\
56 & 2015-Jan-23 & 1132 & -    & 1.01 & -\\
57 & 2015-Jan-28 & 1137 & 0.90 & - & -\\	
\phantom{a}57b & 2015-Jan-29 & 1138 & - & - & 0.84\\	
58 & 2015-Jan-30 & 1139 & -    & 0.96 & -\\
59 & 2015-Jan-31 & 1140 & 0.73 & - & -\\
\phantom{a}59b & 2015-Feb-13 & 1153 & -   & - & 0.91\\
60 & 2015-Feb-14 & 1154 & -    & 1.00 & -\\
61 & 2015-Feb-15 & 1155 & 0.70 & - & -\\	
62 & 2015-Mar-10 & 1178 & 0.80 & - & -\\	
63 & 2015-Mar-13 & 1181 & -    & 1.15 & -\\
64 & 2015-Mar-14 & 1182 & 0.84 & - & -\\	
65 & 2015-Mar-19 & 1187 & 1.00 & - & -\\	
\phantom{a}65b & 2015-Mar-20 & 1188 & - & - & 1.00\\	
66 & 2015-Mar-22 & 1190 & -    & 1.29 & -\\
& & & & \\
\bottomrule
\end{tabular}
\vspace{5mm}
\end{minipage} 
\footnotesize{\textbf{Notes.} The table reports the visit ID, date, time in days from the first observation, presence (via the corresponding seeing value) or absence (-) of a visit in the $r$, $g$, and $i$ bands. The three values in bold identify the selected reference images for each band.}
\end{table*}

\clearpage
\section{Best hyperparameters obtained per classifier} \label{appendixB}
Our fine-tuning process focused on five hyperparameters, shortly described in what follows. \textit{n\_estimators} sets the number of decision trees making up the forest, aiming to balance classification accuracy and computational time. \textit{min\_samples\_split} defines the minimum number of objects required to split a node. \textit{max\_depth} sets a limit on tree depth, aiming to prevent underfitting or overfitting. \textit{min\_samples\_leaf} sets the minimum number of objects for a node to survive, helping control tree size. \textit{max\_features} determines the number of features to take into account when splitting a node.
\begin{table}[!htbp]
\caption{\footnotesize{Set of best values obtained from a grid search-based optimization of the five selected hyperparameters, for each RF classifier tested in this work.}} 
\label{tab:hp}  
\renewcommand{\arraystretch}{1.5}
    \resizebox{\columnwidth}{!}{
\begin{tabular}{l c c c c c c}
\toprule
\ RF classifier & n\_estimators & min\_samples\_split & min\_sample\_leaf & max\_depth & max\_features\\
\midrule
\ $g$, spec-MIR LS & 300 & 10 & 4 & 10 &sqrt\\
\ $r$, spec-MIR LS & 100 & 2 & 4 & 10 & sqrt\\
\ $i$, spec-MIR LS & 500 & 10 & 4 & 10 & sqrt\\ 
\midrule
\ $g$, main LS & 300 & 10 & 4 & 10 & sqrt\\ 
\ $r$, main LS & 300 & 10 & 4 & 10 & sqrt\\ 
\ $i$, main LS & 100 & 10 & 4 & 10 & sqrt\\
\midrule
\ $g$, var. features only & 300 & 2 & 4 & 10 & log2 \\
\ $r$, var. features only & 500 & 10 & 2 & 30 & sqrt\\
\ $i$, var. features only & 100 & 10 & 2 & 10 & log2\\
\midrule
\ \emph{uBrizy}-derived colors only & 300 & 10 & 4 & 10 & sqrt\\
\ all colors only  & 100 & 10 & 2 & 20 & sqrt\\
\midrule
\ $g$, 13 features & 100 & 10 & 4 & 20 & sqrt\\ 
\ $r$, 14 features & 100 & 10 & 4 & 20 & sqrt\\ 
\ $i$, 10 features & 500 & 2 & 4 & 10 & sqrt\\ 
\bottomrule
\end{tabular}
 }
\\
\footnotesize{\textbf{Notes.} The chosen hyperparameters to optimize (see Section \ref{section:classification}) are: \emph{n\_estimators} (tested values: 100, 300, 500), \emph{min\_samples\_split} (tested values: 2, 5, 10), \emph{min\_samples\_leaf} (tested values: 1, 2, 4), \emph{max\_depth} (tested values: 10, 20, 30, None), and \emph{max\_features} (tested options: \textit{sqrt} and \textit{log2}). The first column lists the classifiers evaluated in this work, with their performance metrics shown in Table~\ref{tab:confusion_matrices}.}
\end{table}

\end{document}